# A general model for dynamic contact angle over full speed regime

Hao Wang[1], Lei Chen[2]

[1]The Laboratory of Heat and Mass Transport at Micro-Nano Scale, College of Engineering,

Peking University, Beijing 100871, China

Email: hwang@coe.pku.edu.cn

[2]School of Sciences, Nanchang University, Jiangxi, 330031, China

Email: kingrayo@163.com

Abstract

The prevailing models for advancing dynamic contact angle are under intensive debates, and the fitting performances are far from satisfying in practice. The present study proposes a model based on the recent understanding of the multi-scale structure and the local friction at the contact line. The model has unprecedented fitting performance for dynamic contact angle over wide spreading speed regime across more than five orders of magnitude. The model also well applies for non-monotonous angle variations which have long been considered as abnormal. The model has three fitting parameters, and two of them are nearly predictable at this stage. The one denoting the multi-scale ratio was nearly constant and the one representing the primary frictional coefficient, has a simple correlation with the liquid bulk viscosity.

Dynamic wetting is ubiquitous in natural and man-made systems. It plays important roles in large scale systems such as deposition,[1] coating,[2] and oil recovery,[3] and in small scale systems such as electronics cooling,[4-6] micro and nano fluidics,[7-8] assembly,[9-10] friction,[11] and various biological processes.[12-14] Despite numerous efforts in decades, however, the precise wetting mechanisms remain vague, especially for partially wetting which lacks precursor films.[15] Various hypotheses and theories have been proposed with the debate lasting for years regarding the contact line movement mechanism and the modeling for predicting the profile and contact angles.[16-19] Table 1 depicts the major features of prevailing models for advancing contact lines during liquid spreading. The apparent dynamic contact angle which can be measured by traditional optical methods, $\theta_D$, is usually defined at a scale of microns from the substrate, and the microscopic contact angle, $\theta_m$, is usually defined at a scale of several nanometers within the substrate. One focus of the debate is about the energy dissipation channel. The hydrodynamic model assumes no friction dissipation at the contact line thus $\theta_m$ is a constant.[15, 20-21] While the MKT fully considers the local dissipation and $\theta_m$ is dependent on the moving speed.[16, 18, 22] Accordingly, the different models predicted different liquid film profiles at the contact line as illustrated in Table 1.

Table. 1. Illustration of prevailing models: (a) The traditional hydrodynamic theory,[15, 20-21] where $\theta_m$ is assumed to be a $U$-independent constant and the profile is concave. (b) The molecular kinetic theory (MKT)[16], where $\theta_m$ is assumed to be equal to $\theta_D$ and $U$ dependent. (c) Estimate based on interface rolling motion.[18, 22] (d) The augmented hydrodynamic theory with Van der Waals forces being taken into account.[23]

| Models | Proposed contact line structure | Key features and assumptions | Energy dissipation |
|---|---|---|---|
| (a) Classical hydrodynamic models [19, 24, 25] | 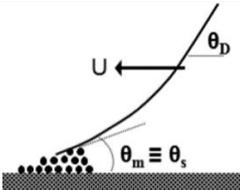 | Two region to establish $\theta_D$; microscopic region truncated; $\theta_D$ changes due to viscous flow. $\theta_m$ serves as boundary condition assumed constant $\theta_m \equiv \theta_{static}$. | Viscous flow in macroscopic region |
| (b) Molecular kinematic theory (MKT) [16] | 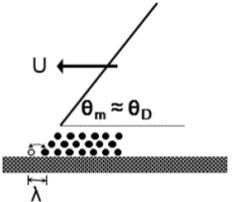 | One region to establish $\theta_D = \theta_m(U)$; $\theta_m$ changes due to molecular displacement at contact line; important parameters include the molecular displacement distance, $\lambda$, and the jump frequency, $\kappa^0$ | Local dissipation at contact line |
| (c) Interface Rolling [18] | 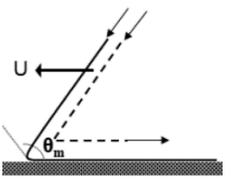 | Two regions to establish $\theta_D$; Interface is rolling and surface tension changes when passing contact line, which changes $\theta_m$. | Both viscous flow and local dissipation |
| (d) Surface Forces (de Gennes') [23] | 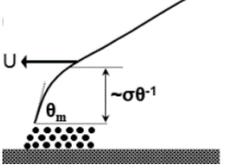 | Two regions to establish $\theta_D$; A bending forms due to long-range intermolecular forces; Bending profile is fixed and $\theta_m$ constant; $\sigma$ is about molecular size. | Viscous flow in macroscopic region |

In practice the variation of the apparent dynamic contact angle, $\theta_D$, with the spreading speed, $U$, is of great interest.[24] In experiments, the $\theta_D$ variation with respect to $U$ could be divided into two groups, i.e., monotonous and non-monotonous. The current models, as wells their combination,[25] have trouble to fit the $\theta_D$ variation over a wide speed range. This is especially true for the non-monotonous variation since all of the current models can only give monotonous increasing of $\theta_D$ with $U$. The hydrodynamic and the MKT as listed in Table 1 are the two most prevailing models. Their fitting performances are as shown in Fig. 1. As shown, the fitting of the MKT or the hydrodynamic model has to be sectional with speed range being no more than two orders of magnitude. Moreover, the fitting totally cannot catch the trend near the middle regime where variation slope comes to be negative. Blake[26] has proposed speculations for the abnormality: For the GC-PET, a change in the wettability of the substrate could occur since gelatin had reaction with water. The angle could be dropping when the speed reaction rate was somehow matching the reaction rate; While for the PET, it didn't react with water, and Blake speculated the surface heterogeneity as the reason. That is, the surface had polar and non-polar cites, and different sites were interacting with water at different speed regimes.

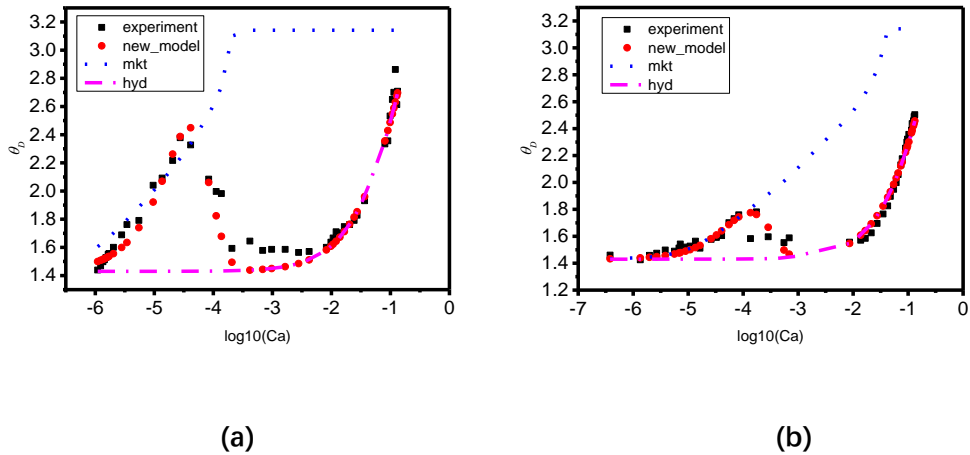

(a)            (b)

FIG. 1. Typical experimental data of non-monotonous variation of the apparent dynamic angle, $\theta_D$, with respect to the advancing speed, $U$. The fitting performances of different models, including the proposed one, are illustrated: (a) Water on GC-PET, reactive.[26] (b) Water on PET, non-reactive.[27]

An important reason for the long-standing puzzle is that the wetting process operates on a scale that extends from the macroscopic to the molecular, while our observations usually involve only macroscopic quantities measured at resolutions no better than several microns.[15] In 2014, Chen *et al.*[28] observed an important mesoscopic structure using a tapping-mode AFM. It was a shoe-tip-like convex nanobending at the advancing contact line. The height scale of the bending was around 20 nm on substrates with sub-nano roughness[28]. The bending profile varied with $U$. The microscopic contact angle, $\theta_m$, extracted at the root of the bending also varied systematically with $U$. Based on the finding Chen et al.[28] has proposed a liquid structure near an advancing contact line. As illustrated in Fig. 2, closest to the substrate was the sublayer region of layers of molecules. The microscopic contact angle, $\theta_m$, is established at the end of this region. Above that it came to the convex nanobending region. The mesoscopic contact angle, $\theta_{me}$, was established at the end of the convex nanobending. Note the nanobending was always convex so $\theta_{me} < \theta_m$, which was important because in this way $\theta_m$ could be great even when the apparent angle was small. In other words, the local angle at the contact line could be much larger than the optically visible contact angle. Especially when $\theta_m > 90°$, the friction reduction due to slippage, rolling, or air entrainment could occur[29-31]. At the end of the convex nanobending it came to the viscous bending region where hydrodynamic analysis was valid. The viscous bending would bend the profile into concave and establish $\theta_D$ which was larger than $\theta_{me}$.

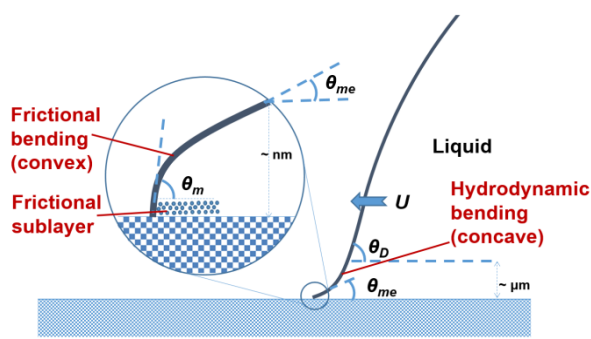

FIG. 2 Schematic diagram of the liquid structure at the advancing contact line of partially wetting liquids. The microscopic contact angle, $\theta_m$, varies with $U$ due to the local friction. The mesoscopic contact angle, $\theta_{me}$, is established at the end of the convex bending and also varies with $U$ due to the local friction.

Most recently, the convex nanobending was successfully reproduced in Liu et al.'s ultra-large-scale molecular dynamics simulation.[32] An important finding was that the convex nanobending would vanish if the substrate was absolutely smooth i.e. frictionless. It was the local dynamic friction form the convex nanobending. The scale of the nanobending would increase with the surface roughness, consistent with their experimental results.[32]

This work proposes a new model to obtain $\theta_D$ based on the above understanding of the contact line structure. Top-downwardly, $\theta_D$ is established after the hydrodynamic viscous bending, therefore according to hydrodynamic theory, [15, 20-21]

$$\theta_D^3 - \theta_{me}^3 = 9\frac{\mu U}{\gamma}A. \quad (1)$$

Here the mesoscopic angle, $\theta_{me}$, rather than $\theta_m$ as in previous hydrodynamic models, [15, 20-21] is used on the left side of the equation.

Then to obtain $\theta_{me}$, we look at the convex bending and the molecular sublayer. We put them together as the local region at the contact line. The friction applies on the local region and deviates the contact angle from the static angle. The force balance is

$$\gamma\left(\cos\theta_s - \cos\theta_{me}\right) = \zeta U, \text{ therefore } \theta_{me} = ac\cos\left(\cos\theta_s - \frac{\zeta U}{\gamma}\right). \quad (2)$$

The friction force is proposed to be calculated by multiplying the mesoscopic friction coefficient, $\zeta$, with the contact line advancing speed, $U$. Note that the convex nanobending is actually a good lubrication structure to minimize the friction at the moving contact line.[32] At higher $U$ when the microscopic contact angle, $\theta_m$, is greater, the nanobending is more like a rolling, rather than sliding on the substrate, which is effective in reduce $\zeta$. Based on this understanding we propose an empirical correlation:

$$\zeta = \zeta_0 \alpha^{U/U_T}, \text{ where } 0<\alpha<1 \text{ and } U_T = \gamma\cos\theta_s/\zeta_0. \quad (3)$$

$\zeta_0$ is defined as the primary frictional coefficient. We use it to represent the frictional coefficient when there is no convex bending and the contact line is simply sliding on the substrate without significant friction reduction like slippage, rolling, or air entrainment, which could be the case for the molecular sublayer at low $U$. According to Eq. (3), $\zeta$ would have a significant drop when $U$ comes to a transition speed $U_T = \gamma\cos\theta_s/\zeta_0$, which corresponds to $U$ that makes a virtual $\theta_m = 90°$ if the sublayer has $\zeta \equiv \zeta_0$. We propose 90° as the transition since after that the liquid is no more a wedge, the shearing of the liquid surface with respect to the liquid-solid interface would be greatly reduced, and also the rolling motion of the convex bending would be much easier; besides, when $\theta >$ 90° the gas domain comes to be a wedge, and air entrapment could happen below the convex nanobending which greatly reduces the friction.

Equations (1)-(3) together predict the apparent contact angle, $\theta_D$. Three fitting parameters are needed at this stage, i.e. $A$, $\alpha$, and $\zeta_0$, while $A$ is nearly a constant, around 10 for different systems. As comparison, the hydrodynamic model needs two fitting parameters and the MKT needs three fitting parameters, and usually the static contact angle has to be fitted with a value that is different from the real value. In the proposed model, the fitting always starts from the system's real static contact angle.

The proposed model has unprecedented fitting performance over wide spreading speed regime, corresponding to $U$ spanning across 5 orders of magnitude. As show in Fig. 1, the abnormal non-monotonous variations are well caught for the first time. At the low-$U$ region, the friction at the contact line dominates and the hydrodynamic viscous bending is relatively weak. The apparent angle increases with the friction increase. As $U$ increases, the frictional coefficient is decreasing and the frictional force will finally start to drop when comes to the middle-$U$ region. If at this stage the hydrodynamic bending is still weak, for example due to low viscosity, the apparent angle has to drop and produce the negative slope. Then comes to the high-$U$ region, the friction is still weak but the hydrodynamic bending is strong enough to push the apparent angle to increase with $U$ again.

According to the model, the monotonous and non-monotonous variations are actually following the same mechanisms. The key is that whether or not the hydrodynamic bending is great enough at the middle-$U$ region. Fig. 3(a) shows a series of results of water-glycerol mixture. The mixtures had similar surface tension while the viscosity increased with the glycerol concentration. It is seen that the hydrodynamic bending is strong enough to avoid the angle dropping when the liquid viscosity is great enough.

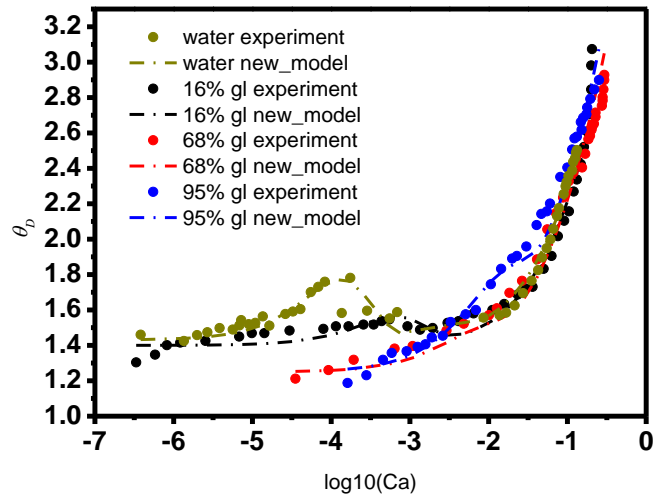

(a)

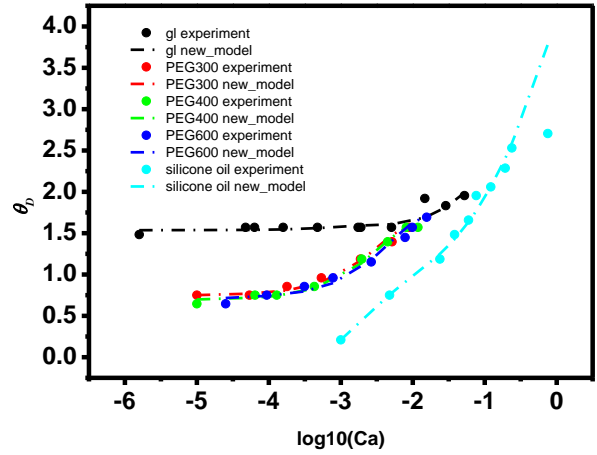

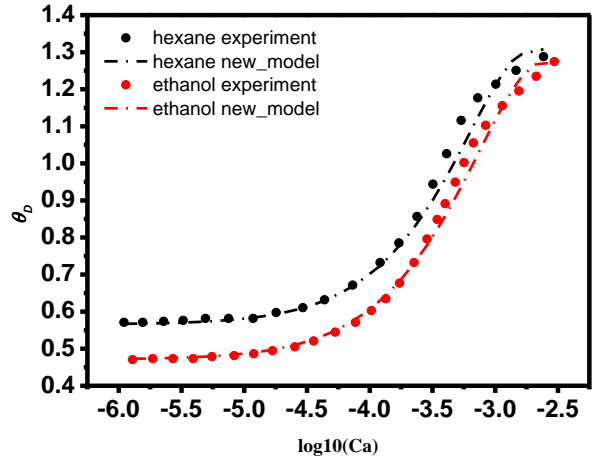

FIG. 3 The fitting performance of the proposed model: (a)A series of results of water-glycerol mixture on PET[27] and the fitting of the proposed model, illustrating that the monotonous and the non-monotonous variations can be unified to the competition between the frictional drop and the hydrodynamic rise. (b) Different liquids on PS.[33] (c) Different liquids on Glass.[34]

Table 2 The properties and fitting parameters for different experiments in the literatures.

| System | Properties | | | Fitting parameters | | |
|---|---|---|---|---|---|---|
| | $\mu$ ($10^{-3}$Pa·s) | $\gamma$ ($10^{-3}$N/m) | $\theta_s$ | $\zeta_m$(Pa·s) | $\alpha$ | $A$ |
| Water, PET[27] | 1.0 | 72.7 | 1.43 | 7.0 | 0.86 | 10 |
| 16% Gl, PET | 1.5 | 69.7 | 1.40 | 1.0 | 0.67 | 9 |
| 68% Gl, PET | 19.0 | 64.8 | 1.25 | 1.2 | 0.45 | 10 |
| 95% Gl, PET | 672.0 | 64.5 | 1.25 | 50 | 0.74 | 12 |
| Gl, PS[33] | 1020.0 | 64.5 | 1.54 | 55 | 0.7 | 9 |
| PEG300, PS | 78.0 | 43.6 | 0.75 | 15 | 0.66 | 10 |
| PEG400, PS | 95.0 | 44.3 | 0.70 | 18 | 0.66 | 10 |
| PEG600, PS | 138.0 | 44.3 | 0.72 | 20 | 0.68 | 10 |
| Silicone oil, PS | 4880.0 | 20.5 | 0.21 | 100 | 0.1 | 8 |
| Hexane, glass[34] | 0.326 | 21.6 | 0.5655 | **0.29** | **0.58** | **10** |
| Ethanol, glass | 1.17 | 19.1 | 0.4707 | **0.63** | **0.55** | **10** |
| | | | | | | |
| water, GC-PET[26] (reactive) | 1.0 | 72.7 | 1.43 | **65.0** | **0.945** | **14** |

Table 2 provides the properties and fitting parameters of the proposed model for experiments in the literatures, corresponding to the fitting performances shown in Fig. 1 and Fig. 3. The values of the three fitting parameters are analyzed as follows. Note the system of water on GC-PET is not included since it is reactive.

First, $A$ is found to be highly stable. The mean value is 9.8 with standard deviation as small as 0.98 for the 11 data points. In hydrodynamic analysis, $A$ is the natural logarithms of the apparent to microscopic scale ratio, i.e., $\ln(L/L_m)$. The value of 9.8 means the scale ratio is around 18000. If the apparent observation is made at microscale, the microscopic scale is then at nano or sub-nano scale, which could possibly smaller than the scale of the convex nanobending. The information indicates that the effective region of the hydrodynamic viscous bending could have a small overlap with the frictional convex nanobending, but the convex bending is still locally dominant since the hydrodynamic bending is achieved over a much larger scale.

The primary frictional coefficient, $\zeta_0$, significantly varies over two orders of magnitude for different systems. Interestingly however, $\zeta_0$ has a close relationship with the liquid bulk viscosity. As shown in Fig. 4(a), the relationship is a simple function,

$$\zeta_0 = 51.68 e^{0.9955 \log_{10} \mu} \quad (4)$$

The adjusted R-squared of the fitting is as good as 0.982. We speculate that the deviation from the fitting curve could reflect the effect of the surface properties such like roughness, which could be a factor in the dynamic friction.[35] The experiments in the literatures didn't provide much data for the surface roughness and we cannot make further analysis at this stage. Anyway, Eq. (4) provides a very useful tool to evaluate $\zeta_0$. In practice we can use the evaluated $\zeta_0$ to predict the trend of the angle variation. For example, the fitting result for water on PET is not very far from the best fitting shown in Fig. 1(b) when using the $\zeta_0$ value predicted by Eq. (4), 2.6, instead of the best fitting, 7.0. A previous research also noticed the relationship between the friction and the viscosity.[36]

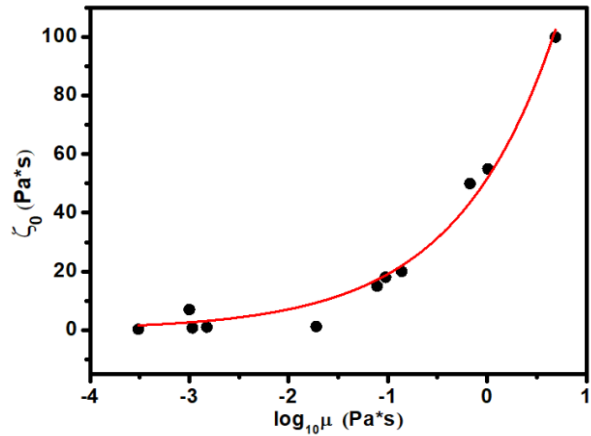

FIG.4 The relationship between the primary frictional coefficient and the liquid bulk viscosity.

$\alpha$ denotes the decay of the local friction at the convex nanobending. It scatters within one order of magnitude. It represents the friction reduction due to possible slippage, rolling, or air entrainment, which could be very complicated for varies systems.

In summary, the proposed model for the first time achieves full-regime fitting. The monotonous and non-monotonous variations are unified. Two out of the three fitting parameters can be well evaluated, which greatly facilitates the practical application.